# Three-terminal field effect devices utilizing thin film vanadium oxide as the channel layer

Dmitry Ruzmetov, Gokul Gopalakrishnan, Changhyun Ko, Venkatesh Narayanamurti, Shriram Ramanathan

*Harvard School of Engineering and Applied Sciences, Harvard University, Cambridge, Massachusetts 02138, USA*

## Abstract

*Electrostatic control of the metal-insulator transition (MIT) in an oxide semiconductor could potentially impact the emerging field of oxide electronics. Vanadium dioxide is of particular interest due to the fact that the MIT happens in the vicinity of room temperature and it is considered to exhibit the Mott transition. We present a detailed account of our experimental investigation into three-terminal field effect transistor-like devices using thin film* $VO_2$ *as the channel layer. The gate is separated from the channel through an insulating gate oxide layer, enabling true probing of the field effect with minimal or no interference from large leakage currents flowing directly from the electrode. The influence of the fabrication of multiple components of the device, including the gate oxide deposition, on the* $VO_2$ *film characteristics is discussed. Further, we discuss the effect of the gate voltage on the device response, point out some of the unusual characteristics including temporal dependence. A reversible unipolar modulation of the channel resistance upon the gate voltage is demonstrated for the first*

*time in optimally engineered devices. The results presented in this work are of relevance towards interpreting gate voltage response in such oxides as well as addressing challenges in advancing gate stack processing for oxide semiconductors.*

## 1. Introduction

Improvements in thin film growth techniques as well as discoveries of striking properties of correlated electron behavior in transition metal oxides such as metal-insulator transition, high temperature superconductivity, and colossal magnetoresistance, result in growing attention to complex oxides from scientific and technological perspectives [1]. The metal-insulator transition (MIT) in vanadium oxide ($VO_2$) has consistently been a focus of intensive experimental and theoretical research. The large magnitude of the resistance change at the MIT, closeness of the transition temperature to room temperature, and fast MIT switching upon optical excitation motivate the interest in MIT in $VO_2$ and suggest this material has high potential for future electronic devices such as switches and logic elements [2, 3]. The origin of the phase transition in $VO_2$ is still under debate [4-6]. In particular, it is not yet clear whether an electric field is or can be the driving mechanism of the MIT. The issue is especially critical for utilizing the MIT effect in electronic applications, since electric, rather than temperature, control of the MIT is highly desirable for fast, reliable operation of electronic devices. At the same time, understanding whether electric field or thermal mechanisms is the primary cause of the phase transition will allow distinguishing between competing theories of the MIT in $VO_2$, specifically whether $VO_2$ is Mott [6] or Peierls [4] insulator.

In conventional observations of the MIT in $VO_2$, the electrical resistance or an optical parameter, such as reflectance, is measured while the temperature of the sample is ramped over the transition temperature ($T_{MIT}$) near 70ºC [7, 8]. A drop in resistance, or a jump in reflectance, is observed once the sample is heated across $T_{MIT}$. It has been also found that an application of sufficiently high voltage may cause the transition to the lower resistance state even if the sample is kept at room temperature [9-13]. In early reports, the MIT induced by a voltage was ascribed to local heating of the $VO_2$ material over the critical temperature $T_{MIT}$ by the current flowing through the device [9]. Recently however, other (non-thermal) causes of the voltage-triggered MIT were proposed [13-15]. Specifically, the electric field, rather than local dissipated power due to Joule heating, was suggested to be the origin of the MIT involving the Mott transition [6, 14, 15] or electrical breakdown [13].

It has proved to be difficult to distinguish whether the temperature or electric field drives the phase transition in experiments on 2-terminal $VO_2$ devices. A voltage applied to a 2-terminal device inevitably provides a current flow that may heat the material over the critical temperature. In the case of homogeneous current flow it has been shown that the heating due to current cannot always account for the observed MIT as concluded based on the electro-thermal simulations that included the heat dissipation through the substrate [16] and based on rough estimates of the transition switching time [11, 14]. However there have been reports of inhomogeneous current flow, such as filamentary conductance, in the low resistance state of $VO_2$ [13, 17]. Formation of local conduction paths, or filaments, allows the possibility of selective heating in excess of the $T_{MIT}$ along the filaments and impedes the study of the process of the MIT with various, say, optical methods that would yield averages over inhomogeneous conducting and insulating areas. For example, the reliable measurement of the temperature inside the conductive filaments is yet to be done, even though some interesting synchrotron micro-spectroscopy experiments visualizing the heat flow in a 2-terminal $VO_2$ device can be noted [18].

The difficulties arising from the current heating when an electric field is applied in a 2-terminal device can be potentially overcome by studying 3-terminal devices similar to a conventional field effect transistor (FET). Realization of 3-terminal field effect devices with $VO_2$ as a channel material proved to be a challenging task. Despite several published reports on the I-V measurements of $VO_2$ in the presence of a gate voltage [14, 19, 20], an explicit and unambiguous account on 3-terminal $VO_2$ experiments still remains an attractive goal. In the published reports so far on the electric field effect in gated $VO_2$ devices [20], there has been little discussion of the physical mechanisms responsible for the observed effects. Also, there is not much detail provided on methods and challenges in fabricating 3-terminal devices that show a non-trivial field effect response.

In this paper we present the results of our experiments and fabrication efforts on three-terminal $VO_2$ devices. We outline the constraints on the device dimensions, experimental challenges in the fabrication of effective $VO_2$ channel and gate insulator, and discuss possible mechanisms of operation of the devices. We observe clear non-trivial response of the channel resistance to the gate voltage in a range of devices. It was found in our experiments that the device fabrication techniques and material synthesis details play a critical role in determining the device response to a gate electric field. Specifically, the quality and fabrication method of $VO_2$/gate dielectric interface profoundly influences the response to the gate voltage by means of the accumulation of the trap charges at the interface and also the magnitude of the MIT in $VO_2$. Our results are important towards utilization of the electrically controlled MIT effect in $VO_2$ in electronic devices.

## 2. Experimental Details

Vanadium dioxide thin films were either reactively DC sputtered in Ar (91.2%) + $O_2$ (8.8%) environment from a V target or RF sputtered in Ar (99.25%) + $O_2$ (0.75%) environment from a $V_2O_5$ target. The $VO_2$ film thicknesses were from 60nm to 150nm. The base pressure in the sputtering chamber was $2x10^{-8}$ Torr and the chamber pressure during sputtering was kept at 10mTorr. The substrate was kept at 550ºC during the deposition. Thin $VO_2$ films synthesized by these sputtering techniques have been comprehensively characterized by a variety of methods [21-23] and the relationships between the morphology of our $VO_2$ films and electron transport and band structure parameters were analyzed in our previous reports [22, 24]. The substrates used were *c*-cut sapphire ($Al_2O_3$) and *n*-type doped Si (001) of resistivity 0.002 – 0.005 Ω cm.

The $VO_2$ film was patterned into a bar with wing contacts for 4-probe resistance measurements. An S1813 resist pattern was produced on top of the $VO_2$ film by photolithography and was used as an etch mask. The dry etching was done in a commercial reactive ion etcher in the Ar(20sccm flow)/$CF_4$(10sccm) gas environment at 90Watt forwarded (40W reflected) power at 60mTorr pressure for 3min. The $SiO_2$ pattern on top of the resulting $VO_2$ pattern was fabricated by means of photolithography and either e-beam evaporation or RF sputtering. Ohmic contacts and the top gate (Fig. 3a, see details below) were deposited during the next photolithography layer, involving thermal evaporation of Au(60 nm, top)/Cr(25 nm), followed by lift-off.

The electrical measurements were performed on a micro-probe station with a temperature-controlled chuck. A programmable current source Keithley 220 established a current through $VO_2$ stripe and a custom built differential electrometer measured a potential difference between two floating (with respect to the ground) voltage terminals. The gate voltage was supplied by a Keithley 230 programmable voltage source or, alternatively, a manually controlled voltage source (for smooth voltage ramps) in the $|V|<120V$ range. Alternatively, the I-V characteristics of 2-terminal devices were measured by a Keithley 236 source meter. Computer-controlled data acquisition from the differential electrometer allowed for the time-monitoring experiments.

## 3. Results and Discussion

Three-terminal devices that have $VO_2$ as the channel material may be considered with the purpose to separate the contributions to MIT due to electric field and Joule heating by the current. The question in mind is whether the MIT can be induced by an electric field without any thermal activation. In an ideal 3-terminal device, the source-drain current would be kept small and well below the value that can cause any heating of the $VO_2$ channel. The gate would induce a static transverse electric field in the $VO_2$ channel that exceeds typical critical electric fields (on the order of $3 \times 10^6$ V/m [11, 12, 15]) causing MIT in 2-terminal devices. If the leakage current through the gate insulator is negligible, then the strong transverse electric field should cause no heating and the presence of the MIT in $VO_2$ would be the proof of the electrostatic origin of the effect.

Despite the apparent simplicity of such an experiment, the MIT effect in $VO_2$ induced by means of a transverse electric field in 3-terminal devices has not been convincingly demonstrated yet.

So, what could be the challenges in conducting experiments with electric field in 3-terminal $VO_2$ devices and what are the device requirements? Let us consider an FET device design consisting of a $VO_2$ thin film channel deposited on an insulating sapphire substrate, source (S) and drain (D) metallic electrodes, and a metallic gate on top of the channel separated from the $VO_2$ by a thin layer of insulator (Fig. 1a). While attempting to fabricate such a device we have found that a high-quality $VO_2$ film sputtered on a sapphire substrate would not conduct if its thickness is below $d_{min}$=45nm. More details of our $VO_2$ synthesis technique and the relationship between film morphology and its electrical properties are described elsewhere [8]. The absence of conductance is attributed to a discontinuous morphology in the $VO_2$ film, consisting of isolated $VO_2$ islands, as confirmed by topographic scans using atomic force microscopy. This tendency has also been observed by other researchers for $VO_2$ films thinner than 50nm [25]. The minimum film thickness is related to the grain size of the $VO_2$ film and depends on the film deposition method. $d_{min}$=45nm corresponds to our film made by reactive DC sputtering from a V target technique. The average grain size of such films was estimated to be 70nm by SEM measurements. The thin films of $VO_2$ with optimal MIT parameters, i.e. exhibiting a sharp 4-order change of resistivity near transition temperature $T_{MIT}$=70ºC, were of ~60nm thickness and thicker and were used as a channel component in our FET devices.

Another aspect of device fabrication is the deposition of an insulator layer on top of the $VO_2$. We have found that RF sputtering of 100nm $SiO_2$ film on top of a 60nm $VO_2$ film reduced the amplitude of the resistance change at MIT by 2 orders of magnitude. The resulting FET devices had a $VO_2$ channel that exhibited a weak 2-order thermal transition near $T_{MIT}$ and were not well suited for the device testing. Deterioration of $VO_2$ material properties may be due to the diffusion of Si into $VO_2$ during $SiO_2$ sputtering although we cannot rule out other mechanisms such as dopant action. Diffusion of Si into the $VO_2$ film after annealing of $VO_2/SiO_2/Si$ hetero-structures was found by Chae *et al* [26] using secondary ion mass spectroscopy (SIMS). In our tests, the initial deposition of $SiO_2$ on top of $VO_2$ at smaller rate resulted in the improvement of MIT strength in the R vs. T curves likely because of the smaller impingement energies of Si ions into $VO_2$ and, consequently, lesser diffusion. The problem was substantially alleviated when e-beam evaporation of $SiO_2$ instead of sputtering was used with properly adjusted deposition parameters

To investigate the influence of the gate insulator on the MIT strength of the $VO_2$ channel we fabricated two identical 2-terminal $VO_2$ devices on the same substrate by means of photolithography and reactive etch. Then an additional $SiO_2$ layer was added to one of the devices on top of the $VO_2$ layer. Figure 2 displays the temperature dependence of the resistance of the two devices. Each of the 2-terminal devices consists of a $VO_2$ stripe (dark thin horizontal stripe in the inset of Fig. 2) on a sapphire substrate contacted with metallic (Au/Cr) leads (white vertical bars on both sides in the inset of Fig. 2). The $VO_2$ stripe extends into large square contact pads (shown partially in Fig. 2) on both sides under metallic leads to provide low resistivity contacts. The large contact pads are necessary to eliminate the distortion of the R vs. T curve above the transition temperature due to the contact resistance as was shown in our previous study [8]. The $VO_2$ stripe size between the gold electrodes is 110μm x 8μm. The $VO_2$ stripe of one of the two devices is covered on top with a rectangular $SiO_2$ pad that is shown in the optical micrograph in the inset of Fig. 2. The data in Fig. 2 show some degradation of the MIT properties of the $VO_2$ channel covered with an oxide layer. Specifically, low T side of the R vs. T curve is lower and high T side is higher in resistance resulting in the less pronounced MIT. In addition, the MIT temperature is shifted to lower T by 3ºC in the stripe under oxide pad. As it was mentioned above, the deterioration of the MIT strength, i.e. ratio of the resistances below and above MIT, $\frac{\boldsymbol{R}(\boldsymbol{T}_{MIT}-\Delta\boldsymbol{T})}{\boldsymbol{R}(\boldsymbol{T}_{MIT}+\Delta\boldsymbol{T})}$, can be over 2 orders of magnitude when the device fabrication process is not optimized. These results emphasize the issue that the material properties inside a multi-component micro-device may deviate from the bulk material properties and therefore the methods of the synthesis of individual components need to be worked out.

In order to achieve a high electric field in the channel material, the thickness of the gate insulator is preferred to be smaller. On the other hand the leakage current needs to be negligible or at least much smaller than the source-drain current. Also, the critical electric field before the electric breakdown of the gate insulator should be as high as possible. The reconciliation of these requirements is challenging when the gate insulator is deposited over a $VO_2$ channel. The generation of defects in the $VO_2$ channel, relatively large surface roughness of the $VO_2$ film, and possible inter-diffusion between the gate dielectric and $VO_2$ result in leakage currents between the gate and $VO_2$ and require a thick gate dielectric layer for such a top-gate geometry. For example, the device in Fig.

3a has 100nm thick $SiO_2$ as the gate insulator. The gate-drain leakage current for this device increases exponentially with the gate-drain voltage ($V_{G-D}$) up to $V_{G-D}$=15V. Above this voltage a permanent damage occurs in the $VO_2$ channel, making it non-conductive. For this reason the measurements on this device were performed for gate voltages $V_G$<10V, when the leakage current was still negligible. Our electrical transport studies of the gate/(gate insulator)/($VO_2$ channel) hetero-structures when the gate/(gate insulator) is either on top of or underneath the $VO_2$ film suggest that the origin of the high leakage current through the gate could be localized defects in the $VO_2$ film that create shorting through the gate dielectric. Smaller in-plane area of the gate, better quality and lower roughness of the $VO_2$ film would decrease the leakage and raise the critical gate voltage. This issue made it necessary to have the gate insulator thicknesses in our experiments at least 100nm and 25nm as grown by e-beam evaporation and atomic layer deposition respectively.

In a recent study, we found that the $VO_2$ thin film free carrier density near room temperature is $2x10^{18}$ $cm^{-3}$ [23]. The 60nm thick $VO_2$ film contains approximately $10^{13}$ electrons per $cm^2$. The changes in the carrier density on the order of $10^{13}$ $cm^{-2}$ can be potentially achieved in a channel due to the gate field before the dielectric breakdown of a typical gate dielectric like $SiO_2$ [27]. The Debye length at 300K in $VO_2$, assuming the carrier density to be $n=2\times10^{18}$ $cm^{-3}$ and dielectric constant $\varepsilon=30$ [28], is on the order of a few nm. Therefore the screening length is significantly smaller than the $VO_2$ film thickness, which complicates the transverse (with no current) field effect experiments in $VO_2$ since most of the material could remain unaffected by the field imposed by the gate. This situation is illustrated in Fig. 1b, where the field lines are drawn in the device. No leakage is assumed through the gate insulator and no Source-Drain current, for simplicity. An application of a positive voltage on the gate will cause the movement of the free carriers (electrons) in $VO_2$ towards the (gate insulator)/$VO_2$ interface. The high concentration of carriers at the interface will lead to the diffusion of the carriers away from the interface that will balance the electrostatic flow. As the result an excess of electrons and *an electric field* will exist inside $VO_2$ film within the screening length ($\lambda_D$) from the interface with the gate insulator. The areas of the charge excess are highlighted in the drawing (Fig. 1b). To consider additional Source-Drain voltage one needs to superimpose horizontal field lines inside $VO_2$ on the drawing in Fig. 1b.

In order for the $VO_2$ channel to be affected throughout its whole volume by the gate field, the $VO_2$ film thickness should be $\lambda_D$=5nm or below. Future experiments would consider the use of ultra-thin channel layers possibly grown by epitaxial / layer-by-layer techniques. Despite the fact that the electric field from the gate can affect only a thin superficial layer of $VO_2$ near the interface with the gate dielectric, there have been a few experiments on gated $VO_2$ devices reporting non-trivial results even for thick (above 60nm) polycrystalline $VO_2$ films [14, 19, 20]. In an early report of Vasiliev *et al.*, a DC gate voltage did not produce any effect on the Source-Drain I-V curves, however the resistance switching in $VO_2$ channel was observed when the gate voltage was applied in pulses [19]. Stefanovich *et al.* observed both permanent and transient switching of the $VO_2$ resistance when the gate voltage was continuously ramped up [14]. In these experiments, the thick gate insulator (70nm of $SiO_2$) and thick $VO_2$ film (with loosely defined thickness in the range 0.1 – 1 μm) preclude any substantial transverse electric field throughout the $VO_2$ channel. The resistance switching was ascribed to the Mott transition due to the carrier injection from the gate into the $VO_2$ channel [14]. In the measurements on 3-terminal devices by Kim *et al*. [20], a $SiO_2$ gate insulator was used and the application of a gate voltage produced shifts of the critical Source-Drain voltages that trigger the MIT. However the shift occurs in different directions when the magnitude of the gate voltage is increased. More experimental data on such devices would be desirable. Summarizing, in the reported so far experiments on 3-terminal $VO_2$ devices some MIT effects due to a gate voltage were observed. The origin of the MIT could be carrier injection, rather than a DC electric field. Further work is hence needed to understand the observed effects.

In the present work a series of $VO_2$ gated devices has been studied. An optical micrograph of device A with added labels and results of measurements on this device are presented in Figure 3. The $VO_2$ stripe shown in Fig. 3a has wing contacts for 4-probe resistance measurements with current going from $I_+$ to $I_-$ and potential difference evaluated between $V_+$ and $V_-$ leads. Thus this device design allowed monitoring the resistance only of the part of the $VO_2$ strip covered entirely by the gate which provided the high sensitivity of the device to the gate field. The 4-probe resistance of the $VO_2$ bar under the gate was measured as a function of temperature for zero gate bias (Fig. 3b). Fig. 3b shows that in the absence of the gate voltage the $VO_2$ bar exhibited an excellent MIT of about 4 orders of magnitude near the $T_{MIT}$=70ºC. Therefore the material quality was not significantly degraded by the employed fabrication technique.

The $VO_2$ film thickness in the device depicted in Fig. 3a is 100nm. There are a few device operation scenarios that can be tested with such a device. First, the electric field from the gate could induce the transition to metallic state in a thin superficial layer in $VO_2$ (highlighted in Fig. 1b) initially at low (below $T_{MIT}$) temperature that would shunt the overall resistance of the $VO_2$ channel. The cross-section, and consequently conductance, of this superficial layer is roughly a factor of 20 smaller than the channel cross-section. Hence, the MIT in the superficial layer by 2 - 3 orders of magnitude in the resistance would be clearly detectable. Another possible mechanism that could lead to a measurable effect caused by the gate voltage is hot carrier injection into $VO_2$ channel. Carrier injection was suggested to be the reason of the field effect in $VO_2$ 3-terminal devices reported so far [14, 20] as well as the reason of the optically detected MIT in photo-excited $VO_2$ thin films [5] and electrically gated devices [29]. Also, importantly, the MIT systems close to the critical transition point may be much more susceptible to the carrier density change than conventional semiconductors. In this regard, it is worthwhile also to note the experiments on colossal magnetoresistance materials [30] where the resistance was shown to change significantly upon gating even though the channel thickness was substantially larger than the Debye length suggesting alternative mechanisms to just electrostatic modulation.

In our experiments, the application of the gate voltage caused non-trivial response of the I-V characteristics for a number of studied devices (that were identical to the Device A shown in Fig. 3a, i.e. channel length 240µm, and also sizes 1200µm, 600µm, 60µm). Figure 3c shows how I-V characteristics of the 240µm device A responded to the application of the gate voltage of -10V when the sample was kept at T=30ºC. The gate voltage was applied between the top gold terminal and metallic chuck on which the device substrate was placed. The I-V curves remained linear upon the gating but the 4-probe resistance (V+, V-, I+, I-) was shifted up. The resistance continued to increase for some time (~10min) even after the gate voltage was removed. Application of a positive gate voltage of up to +10V did not seem to affect the increased resistance. The gate voltages of 15V or above apparently caused the break-down of the gate dielectric and damaged the $VO_2$ material. So that after the break-down, the 4-probe resistance either strongly increased and did not display MIT in R vs. T curves anymore or corresponded to an open circuit. The inset in the figure 3c describes how the resistance of $VO_2$ under the gate changes with time when a −10V potential is applied to the gate. The resistance (equal to $V_+ - V_- / 100\,\mu A$) monotonously increases with time. The measurement was stopped after 15min because prolonged exposure to the gate voltage could permanently damage the sample.

Larger samples (1200μm and 600μm channel length compared to Device A) showed higher leakage currents between the gate and the $VO_2$ channel and had correspondingly smaller if any response to the gate voltage. Whereas the devices with smaller channel length showed larger response to the gate voltage but the response was often non-reproducible and depended strongly on the history of the measurements. Fig. 4 presents an example of the electrical testing of the top gate sample (Device B) similar to the Device A (Fig. 3a) but with the smaller channel length of 60μm. One can see in Fig. 4 that an application of a voltage to the top gate separated by a $SiO_2$ layer from the $VO_2$ channel may modulate the channel resistance by almost 5%. A common trend seen in the electrical response of both samples A and B to the gate voltage is that the resistance change continues increasing with time, even after removal of the gate field. However, the magnitude of the effect depends on the electrical measurement history and thermal history of the device. Devices A and B were fabricated and tested one year apart and differ only in size and in some minor details of $VO_2$ film synthesis (both were synthesized by reactive DC sputtering). The temporal behavior in figures 4 and 3c, sample-to-sample variation, and measurement history dependence indicate that the slow filling of traps at the $VO_2$/(gate insulator) interface may be a possible cause of the observed gate effect. The annealing of the sample could be considered to alleviate this problem. However annealing at temperatures of 150ºC and above deteriorates the MIT parameters of the $VO_2$ due to transformation of $VO_2$ into other $VO_x$ ($x \neq 2$) phases and also due to inter-diffusion of the material at the interfaces. The problem, noted above, of an apparent "gated field effect" on the $VO_2$ channel is therefore an important issue.

As discussed here, the quality of the gate dielectric layer and its interface with $VO_2$ are critical factors for detecting the field effect in 3-terminal $VO_2$ devices. For this reason we have explored a number of ways of synthesizing the gate dielectric within the device: e-beam evaporation of $SiO_2$ and $Al_2O_3$, RF sputtering of $SiO_2$ and $Si_3N_4$, and atomic layer deposition of $HfO_2$ and $Al_2O_3$. It was found for the gate-on-top-type devices (e.g. Fig 3a) that, while the high quality of $VO_2$ can be preserved within the hetero-structure, the devices suffer from large leakage through the gate dielectric and poor dielectric/$VO_2$ interfaces (e.g. with trapped charges). However, we found that the devices with the gate below the $VO_2$ film offer better quality of the gate dielectric layer and its interface as opposed to the gate-on-top devices.

The difficulties associated with using the bottom gate in $VO_2$ devices are the following. The deposition of a $VO_2$ film on top of the (gate substrate)/(dielectric film) base results in degradation of MIT parameters of $VO_2$, since $VO_2$ grows better on single crystal substrates. Also, our initial experiments with the growth of $VO_2$ films on top of metallic gates (such as Au etc.) produced non-uniform $VO_2$ films or clusters of $VO_2$. The latter problem can be overcome by using a heavily doped semiconducting substrate as a conducting bottom gate (instead of a conventional metal film or substrate). Notably, doped semiconductors or ceramics were used as a bottom gate in the few published reports on 3-terminal $VO_2$ devices so far [14, 20]. Conducting doped-Si substrates allow synthesis of smooth $VO_2$ films, but the quality of $VO_2$ degrades when it is grown on a hetero-layered structure. This is illustrated in Fig. 5 where the $VO_2$ resistance curves are shown for the $VO_2$ film deposited simultaneously by means of RF sputtering from $V_2O_5$ target on single crystal $Al_2O_3$ substrate and *n*-doped Si substrate covered with highly uniform $Al_2O_3$ film (25nm thick) grown by atomic layer deposition (ALD). One can see that the deposition of $VO_2$ on ALD-$Al_2O_3$/*n*-Si decreases the magnitude of MIT by one order as compared to the sapphire substrate. The fabrication of the top gate and electrodes in the device further degrades the MIT probably due to inter-diffusion of the Si from the gate dielectric into $VO_2$.

In the device discussed below, top and bottom gates were fabricated simultaneously (Device C). A highly uniform $Al_2O_3$ insulating layer was deposited by ALD on top of *n*-Si conducting substrate and served as a base for $VO_2$ growth by means of RF sputtering from $V_2O_5$ target. The ALD-grown gate insulator/$VO_2$ interfaces exhibited reproducible electrical response to applied gate voltages, with no time dependence (cf. Fig 4) or persistence beyond removal of the gate voltage. We describe below the experiments with bottom gate biasing while no voltage is applied to the top gate.

Gate modulation of the channel resistance in Device C was investigated by repeating a schedule consisting of three successive I-V measurements, performed using a Keithley 236 Source Measure Unit, with each measurement separated in time by 15s. The first and last measurements were performed with the gate voltage switched off, while the second measurement was performed with the gate on. The difference ΔR between the "on" resistance and the mean value of the two "off" resistances is displayed in Fig. 6, where error bars represent standard deviations calculated from dozens of measurements performed at T=60ºC. While we detect a systematic decrease in resistance for a negative applied gate voltage, *no change is detected for similar positive voltages*. Due to sharply increasing leakage through the gate oxide at higher gate voltages, we were restricted to gate voltages of 0.5V magnitude. At these small gate voltages, a drop in the channel resistance of about 120Ω , corresponding to a change by −0.26% (± 0.04%) is measured for a negative bias. Furthermore, while the above effect was observed in this device at 60ºC and 65ºC, there was no modulation in the channel resistance detected for either polarity of gate voltage when the device was characterized at lower temperatures.

The temperature and polarity dependence of the observed gate modulated channel resistance rule out the possibility that this effect is caused by the leakage current from the gate, which is measured to be smaller than 1% of the typical S-D current flowing through the channel during a voltage sweep. Moreover, such a polarity dependence is expected from the Mott picture, in which a critical carrier density induces a transition to metallic behavior. Due to spatial inhomogeneities in the carrier concentration, the injection of charge into the channel, already near the critical density at T=60ºC, advances the nucleation of metallic domains within regions of higher charge density due to the early realization of a critical carrier density within these regions. This carrier induced formation of the metallic domains within a semiconducting matrix results in an enhancement of its average conductivity. On the other hand, removal of charge by reversing the gate polarity has no effect on the transition, and affects the resistance of the semiconducting channel much more weakly via the relative change in the total carrier concentration.

Application of a negative bias on Gate 2 (see Fig. 6 inset) in Device C creates extra positive charges in $VO_2$ at the $VO_2/Al_2O_3$ interface (in a manner analogous to the illustration in Fig. 1b). Thus, the polarity of the gate effect observed in our experiment is consistent with results published in literature. In particular, the optically observed effect of the gate was ascribed to the accumulation of holes at $VO_2$/(gate insulator) interface due to a negative gate bias in experiments of Qazilbash *et al*. [29]. The metal-insulator transition in $VO_2$ was argued by Kim *et al.* to be due to the Mott mechanism and triggered by the hole carrier doping of the valence band [6]. The transition of $VO_2$ to the metallic phase was also observed by means of photo-excitation resulting in the hole doping of the valence band, but $VO_2$ was argued to be a band insulator judging by the time scale of the MIT [5]. Therefore the polarity of the gate effect is consistent with the published results.

The significance of the results of the gate effect experiments summarized in Fig. 6 is that we observe for the first time the modulation of the channel resistance due to the gate voltage in all-electrical (the switching is induced and detected electrically) three-terminal $VO_2$ devices with no hysteresis. An application of a negative gate potential reduces the channel resistance and the removal of the gate potential brings the channel resistance to the initial state with no detectable time delay (i.e. less than 1sec). The gate effect described in Fig. 6 is highly reproducible within the error bars and was tested for dozens of gate voltage cycles. The reversible control of the $VO_2$ resistance in our experiments represents an important step forward towards utilizing the MIT effect in $VO_2$ in future electronics. The application of larger gate voltages on this device was restricted by the presence of leakage currents through the gate dielectric. Further optimization of the gate insulator growth and device design is necessary in order to increase the magnitude of the gate effect and characterize the dependence of the current-voltage characteristics of the channel on the gate voltage. The improvement of the $VO_2$ synthesis on top of substrate/(gate insulator) hetero-structures and, consequently, the improvement of MIT parameters of the $VO_2$ in bottom gate devices may be another way of increasing the magnitude of the gate effect. Indeed, the resistance change at the MIT in $VO_2$ on single crystal substrate in Fig. 3b is by two orders of magnitude larger than in $VO_2$ of the bottom gate device (Fig. 5).

## 4. Conclusions

We investigated in detail the aspects of fabrication and electrical measurements of three-terminal devices utilizing $VO_2$ as a channel material. The gate dielectric and its interface with $VO_2$ plays a crucial role in the performance of the devices. Poor quality dielectrics or dielectric/ $VO_2$ interfaces may result in an apparent field effect of the gate on the $VO_2$ channel resistance that is time dependent and not reversible. Such field effects may not be associated with the metal-insulator transition in $VO_2$ and need to be carefully considered.

On the other hand, optimally engineered gate/gate dielectric/$VO_2$ hetero-structures are demonstrated to exhibit reliable switching of the channel resistance upon application of a gate voltage. The switching is reproducible for dozens of cycles, reversible (i.e. the resistance returns to the initial state after removal of the gate voltage), and manifests only for negative polarity of the gate voltage at temperatures close to the MIT transition. The latter observations indicate that the reported effect likely occurs due to the metal-insulator transition induced by the hole injection into a thin region of the $VO_2$ channel along the interface with the gate dielectric. These results are of relevance towards utilizing the MIT effect in $VO_2$ in future electronic devices. Subsequent work on such devices will be aimed at exploring different dielectrics and growth conditions towards improving leakage characteristics and thereby possibly enhancing the field effect through application of larger gate voltages.

## Acknowledgements

We acknowledge NSF supplement PHY-0601184 and AFOSR for financial support. Device fabrication was performed, in part, at the Harvard University Center for Nanoscale Systems (CNS), a member of the National Nanotechnology Infrastructure Network (NNIN) which is supported by NSF Award No. ECS-0335765.

Figures.

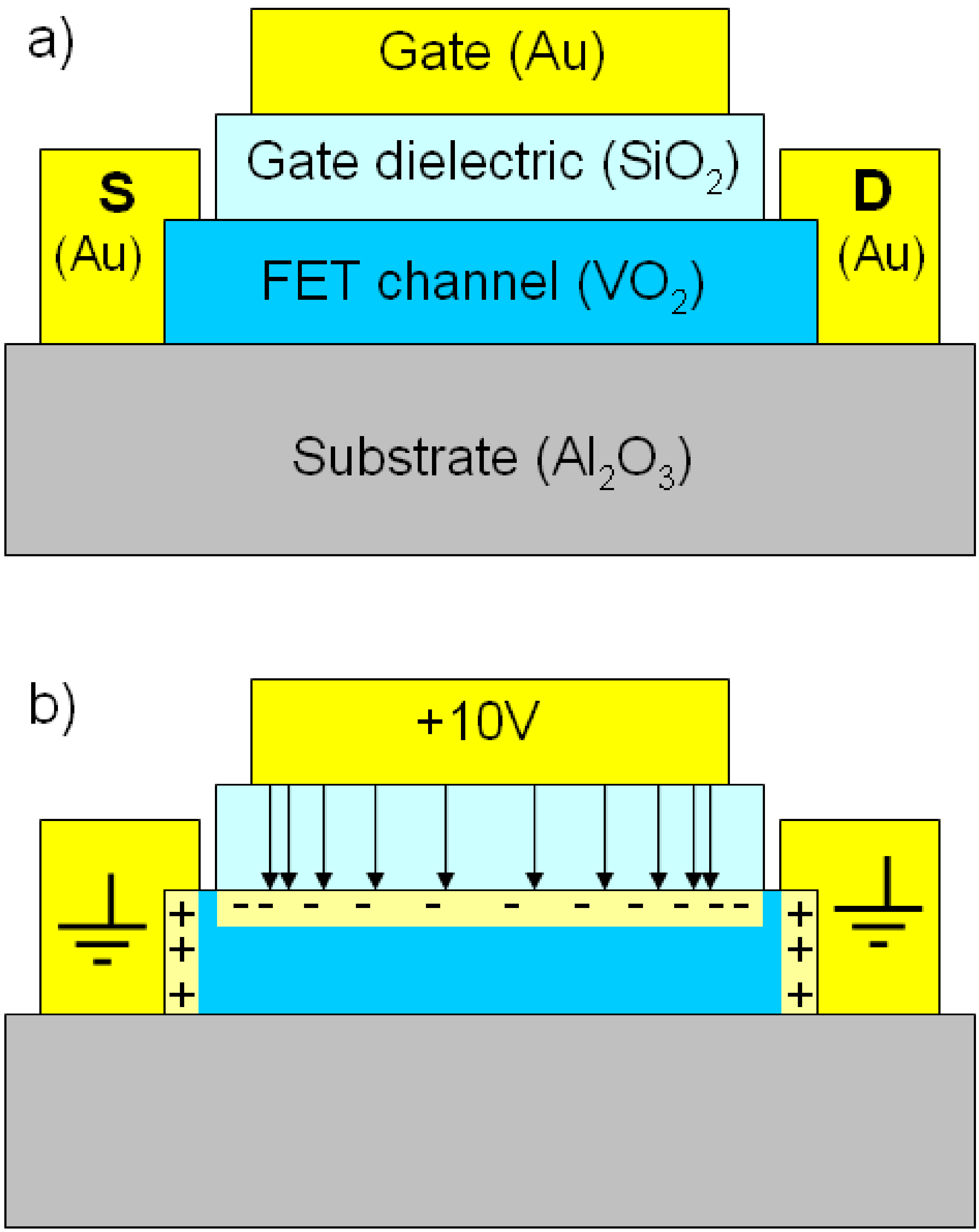

**Figure 1.** A diagram of a 3-terminal device. **a)** Notations. The current between the Source (S) and Drain (D) is modulated by the potential on the Gate separated from the $VO_2$ channel by a Gate insulator layer; **b)** Application of a positive voltage on the gate will create a field in the gate dielectric that penetrates into the $VO_2$ channel likely within the screening length $\lambda_D$. An additional small Source-Drain voltage would superimpose horizontal field lines in the channel.

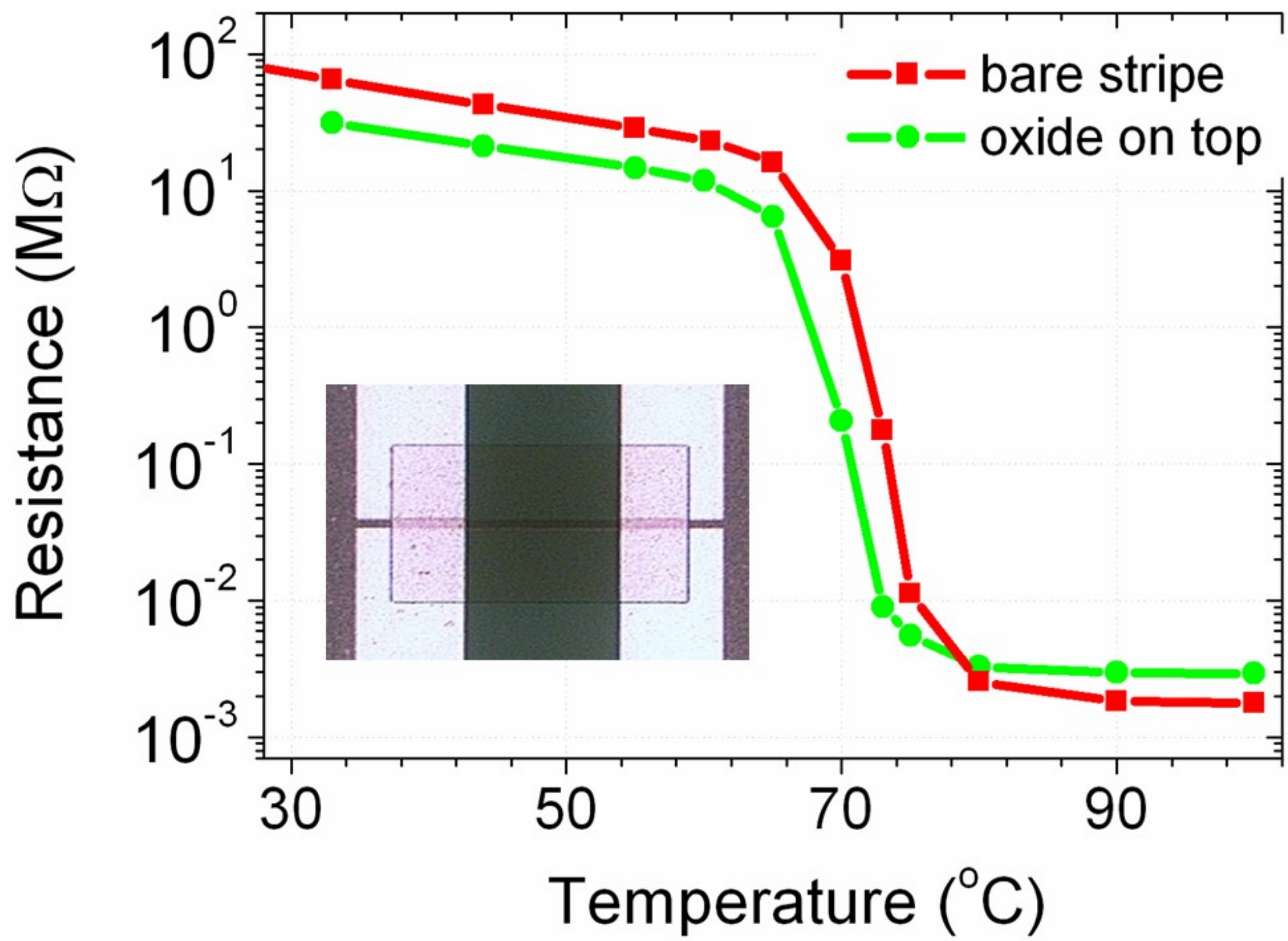

**Figure 2.** Results of electrical testing of two photo-lithographically patterned $VO_2$ channels of identical geometries. Each $VO_2$ channel was deposited on a sapphire substrate, patterned into a stripe shape, and contacted with gold pads. One of the $VO_2$ channels is covered with a $SiO_2$ thin film pad that affected its MIT properties. **Inset:** image of the $VO_2$ device with $SiO_2$ pad on top.

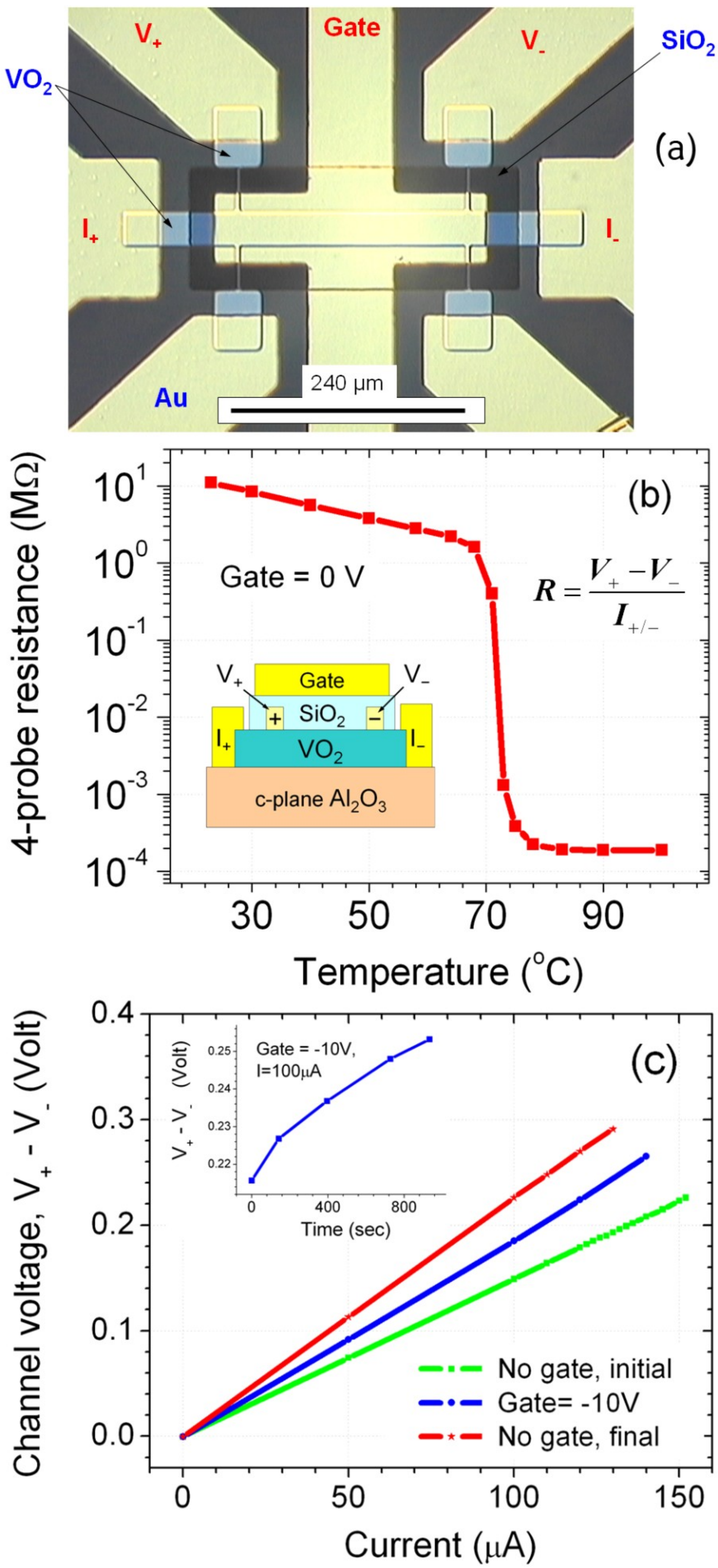
V+
Gate
V-
SiO2
VO2
(a)
I+
I-
Au
240 µm
4-probe resistance (MΩ)
(b)
Gate = 0 V
R = (V+ − V−) / I+/−
Gate
V+
V−
SiO2
I+
VO2
I−
c-plane Al2O3
Temperature (°C)
Channel voltage, V+ - V- (Volt)
(c)
Gate = -10V,
I=100µA
V+ - V- (Volt)
Time (sec)
No gate, initial
Gate= -10V
No gate, final
Current (µA)

**Figure 3.** Gate-Source-Drain $VO_2$ device of 240μm channel length (Device A). **a)** An optical micrograph of the device with labeled elements. **b)** MIT in the 4-probe ($I_+$, $I_-$, $V_+$, $V_-$) resistance of the device with no voltage on the gate. **Inset:** Device diagram; **c)** I-V curves as a function of gate voltage at T=30ºC. **Inset:** Effect of the gate voltage on the $VO_2$ channel as the function of time at 30ºC.

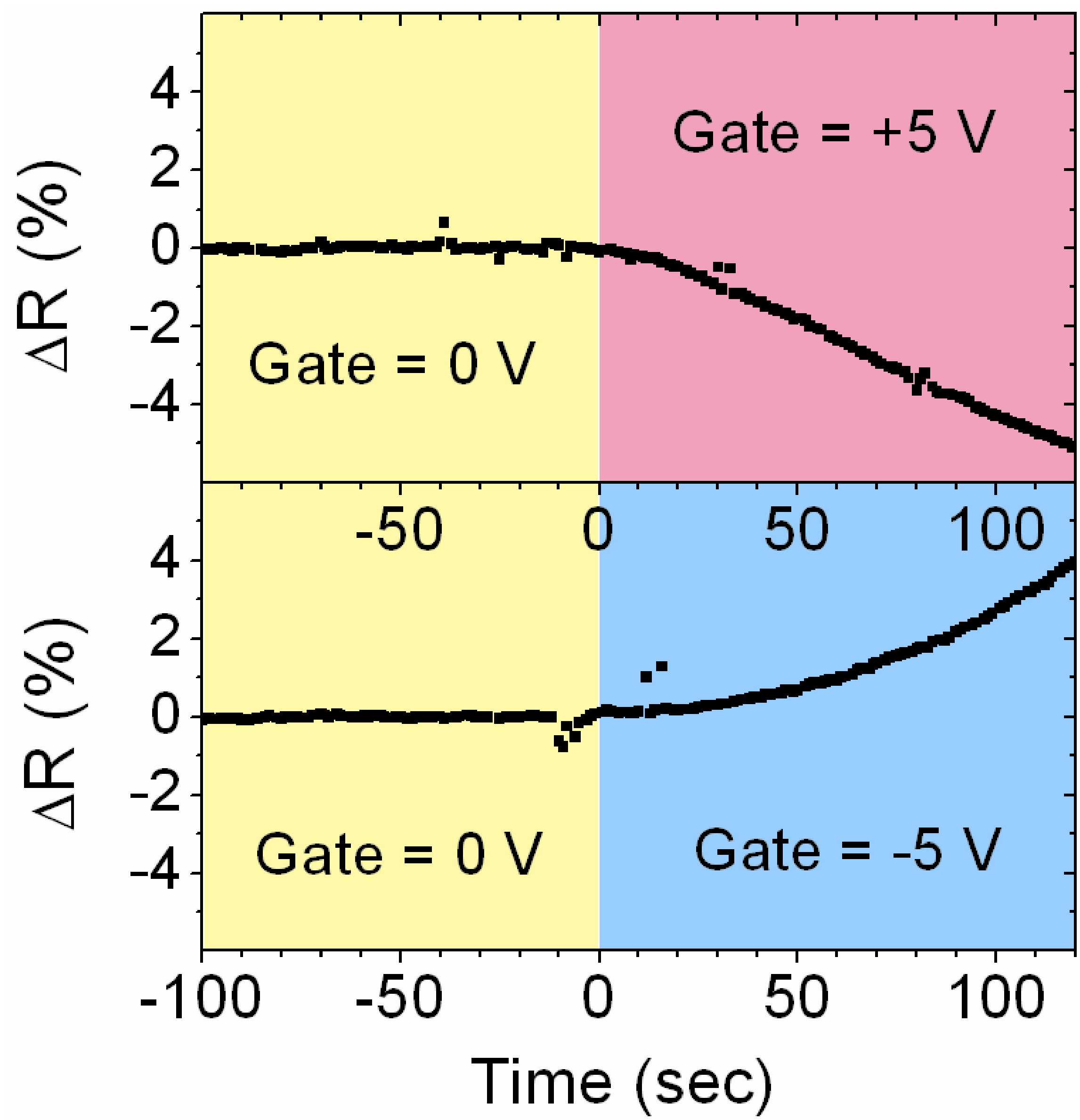

**Figure 4.** Gate-Source-Drain $VO_2$ device of 60μm size (Device B). The effect of the gate voltage as a function of time in a selected experiment. The change of the channel resistance due to positive (top panel) and negative (bottom panel) gate voltage.

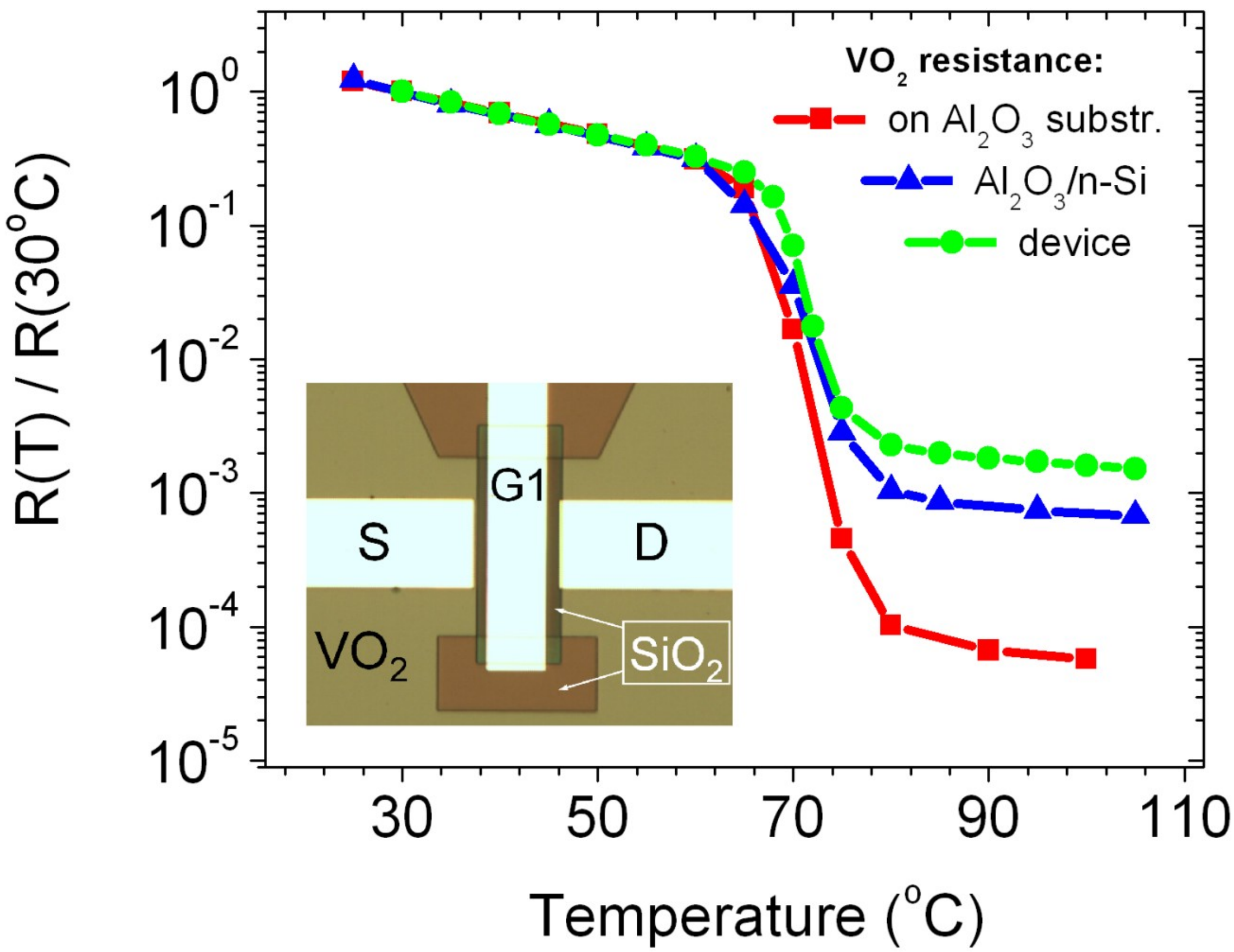

**Figure 5.** Normalized resistance of $VO_2$ film upon fabrication of a 3-terminal device on a conductive n-Si substrate used as an electric gate (Device C). The resistance curves are shown for a $VO_2$ film grown on a *c*-cut $Al_2O_3$ substrate; on a *n*-Si substrate capped with ALD-grown $Al_2O_3$ layer; and source-drain resistance of the fabricated device with no voltage on the top (G1) and bottom (G2) gates. **Inset:** A micrograph of the device. Bright squares (S, D, G1) are gold contacts. The source-drain separation is 110μm.

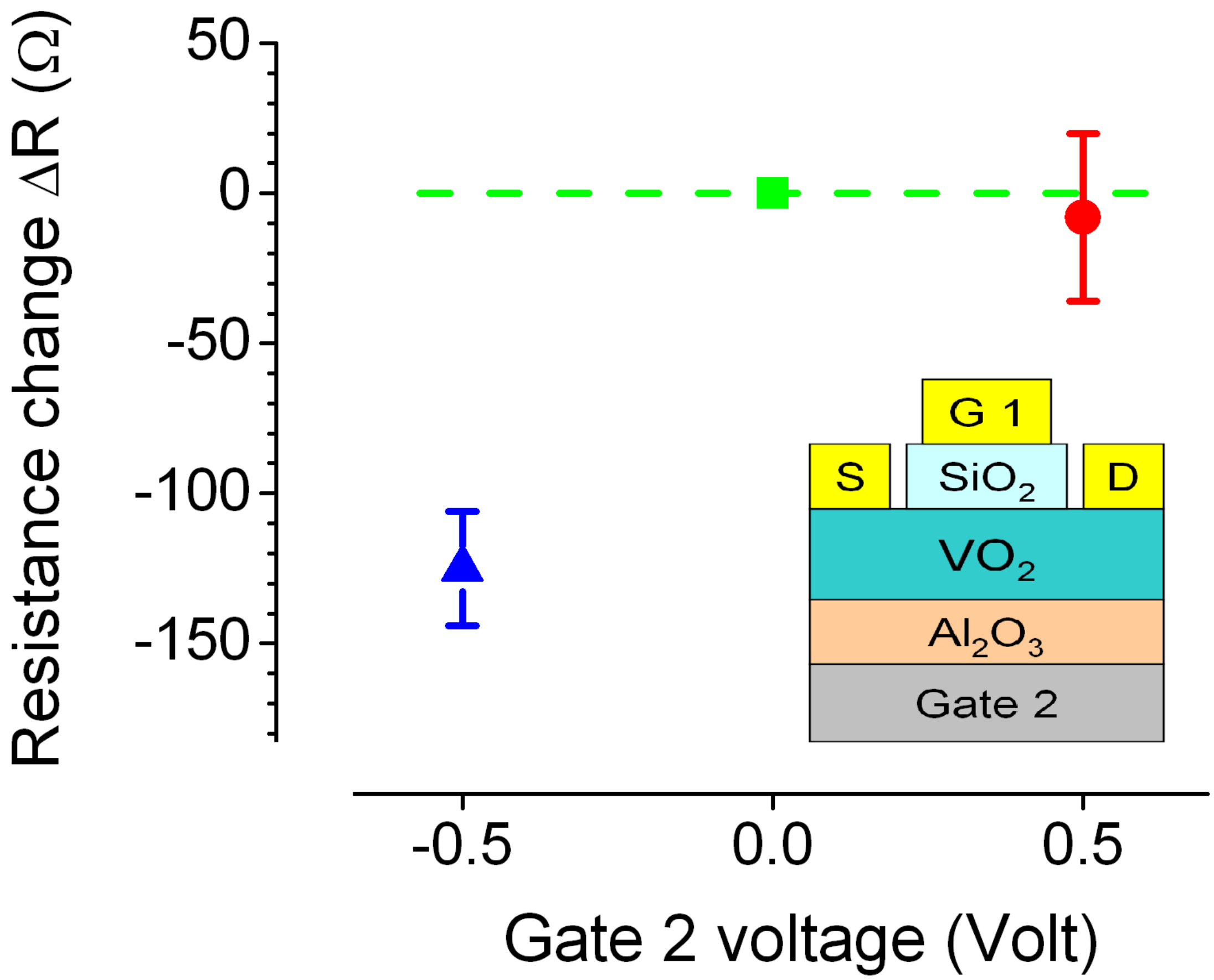

**Figure 6.** The effect of the gate voltage on the source-drain resistance of the $VO_2$ channel in Device C. The application of the negative voltage to the bottom gate 2 decreases the resistance, while positive voltage produces no noticeable effect. The data are well reproducible within the error bars. **Inset**: the diagram of the device.